\documentclass[12pt]{iopart}

\usepackage[english]{babel}
\usepackage{enumerate}
\usepackage{multicol}
\usepackage{fancyhdr}
\usepackage{amssymb}
\usepackage{hhline}
\usepackage{wrapfig}
\usepackage{caption}
\usepackage[dvips]{graphicx}
\usepackage{pifont}
\usepackage{makeidx}
\usepackage{cite}
\usepackage[T1]{fontenc}

\newcommand{\lco}{LaCoO$_3$}
\newcommand{\lsco}{La$_{2-x}$Sr$_x$CoO$_4$}

\newcommand{\tg}{$t_{2g}$}
\newcommand{\eg}{$e_{g}$}
\newcommand{\LS}{$t_{2g}^{\,6}e_{g}^0$}
\newcommand{\IS}{$t_{2g}^{\,5}e_{g}^1$}
\newcommand{\HS}{$t_{2g}^{\,4}e_{g}^2$}

\newcommand{\ca}{$\chi_{ab}$}
\newcommand{\cc}{$\chi_c$}
\newcommand{\ch}{susceptibility}
\newcommand{\cd}{$c$ direction}

\begin{document}

\title{Anisotropic Susceptibility of La$_{2-x}$Sr$_x$CoO$_4$ related to the Spin States of Cobalt}

\author{N~Hollmann$^1$,
M~W~Haverkort$^1$,
M~Cwik$^1$,
M~Benomar$^1$,
M~Reuther$^1$,
A~Tanaka$^2$,
T~Lorenz$^1$}
\address{$^1$II. Physikalisches Institut, University of Cologne, Z\"{u}lpicherstrasse 77, D-50937 K\"{o}ln, Germany}
\address{$^2$Department of Quantum Matter, ADSM, Hiroshima University, Higashi-Hiroshima 739-8530, Japan}

\ead{hollmann@ph2.uni-koeln.de}

\begin{abstract}

We present a study of the magnetic \ch\ of \lsco\ single crystals
in a doping range $0.3\leq x \leq 0.8$. Our data shows a
pronounced magnetic anisotropy for all compounds. This anisotropy
is in agreement with a low-spin ground state ($S=0$) of Co$^{3+}$
for $x\geq 0.4$ and a high-spin ground state ($S=3/2$) of Co$^{2+}$.
We compare our data with a crystal-field model calculation
assuming local moments and find a good description of the magnetic
behavior for $x\ge 0.5$. This includes the pronounced kinks observed in the inverse
magnetic susceptibility, which result from the
anisotropy and low-energy excited states of Co$^{2+}$ and are not
related to magnetic ordering or temperature-dependent spin-state transitions.

\end{abstract}


\maketitle

Transition-metal oxides are known for their complex interplay
between different degrees of freedom like spin, charge and
orbitals. In some of these systems another interesting property
is found: ions like Co$^{3+}$ in a crystal-field environment can
occur in different spin states. Additionally, transitions between
different spin states are possible for some compounds. A prominent
example showing this phenomenon is \lco, which has been examined
and discussed since the middle of the last century
(see e.g. Refs.~\cite{jonker53a,racah67a,abbate93a,korotin96a,
zobel02a,maris03a,noguchi02a,ropka03a,lengsdorf04a,baier05a,haverkort06a}).

The existence of different spin states arises from the competition
between crystal-field effects and on-site Coulomb interaction. Crystal
fields lift the degeneracy of the $3d$ states. If the crystal
field is strong, this can lead to a violation of Hund's rules. In the case
of cubic symmetry and in a one-electron picture, the five-fold
degenerate $3d$ states are split into a three-fold degenerate \tg\
and a two-fold degenerate \eg\ level. The splitting between \tg\
and \eg\ states is called $10Dq$. With a strong crystal field, the electrons will be forced
into the low-spin state (LS). Regarding a $3d^6$ system, this state
consists of an antiparallel alignment of spins with \LS\ and
$S=0$. On the other hand, the Coulomb interaction manifests itself
in the effect of Hund's coupling. A parallel arrangement of spins
minimizes the electron-electron repulsion because the Pauli
principle forces the electrons to occupy different orbitals. In a
weak cubic crystal field, this effect will dominate and lead to
the high-spin state (HS), which is the configuration with the
highest total spin possible in accordance with the Pauli
principle. For a $3d^6$ system, this configuration is \HS\ with
$S=2$. For Co$^{3+}$ the crossover between these two different
spin states occurs roughly at an energy difference of
$10Dq=2.2$eV.

In the case of LaCoO$_3$, a low-spin ground state for Co$^{3+}$
was found\cite{racah67a}. Interestingly, the difference between
the crystal field energies and the promotional energies is so
small that an excited state with different spin state can be
reached by thermal excitation. This excited state has been a subject
of debate for a long time. Apart from the HS state described above, the intermediate-spin state (IS,
\IS\ $S=1$) was also discussed\cite{korotin96a} and reported in many experiments,
but it was shown that it might have been confused
with a spin-orbit coupled HS state\cite{haverkort06a}.

For the layered cobaltates \lsco, much less
is known about the spin state of Co$^{3+}$. Based on magnetic
measurements\cite{moritomo97a} a HS
ground state for $x\leq 0.7$ and a spin-state transition
to an IS ground state for $x>0.7$ was proposed. This conclusion was based on
a Curie-Weiss analysis of the
\ch\ in a temperature range of 100K to 300K and NMR measurements\cite{itoh99a}. Unrestricted Hartree-Fock
calculations\cite{wang00a} showed a slightly different picture. Here, three
different magnetic phases were found. An antiferromagnetic HS
phase ($x< 0.39$), a ferromagnetic HS phase ($0.39\leq x\leq 0.52$) and
an antiferromagnetic LS-HS-ordered phase ($x>0.52$) were proposed. These calculations
were also based on the results of the Curie-Weiss analysis in Ref.~\cite{moritomo97a}.
Taking into account that \lsco\ is an anisotropic material with
rather strong spin-orbit coupling compared to the tetragonal crystal field splitting, the validity of the Curie-Weiss
law is questionable. The aim of this paper is to analyze the magnetic \ch\ from a different perspective,
concentrating on the spin state of Co$^{3+}$.

The single crystals used for the magnetic measurements have been
grown using the floating-zone technique in an image furnace. A
strontium doping range of $0.3\leq x\leq 0.8$ was covered.
Resistivity measurements revealed that \lsco\ is a strong
insulator for all Sr doping concentrations. The $x=0.5$ sample
turned out to possess the highest resistivity, which is in
accordance with the charge ordering of Co$^{2+}$ and Co$^{3+}$ at
$\approx 750$K that has already been reported\cite{zaliznyak00a}.

\begin{figure}[t]
\includegraphics[angle=0,width=\textwidth]{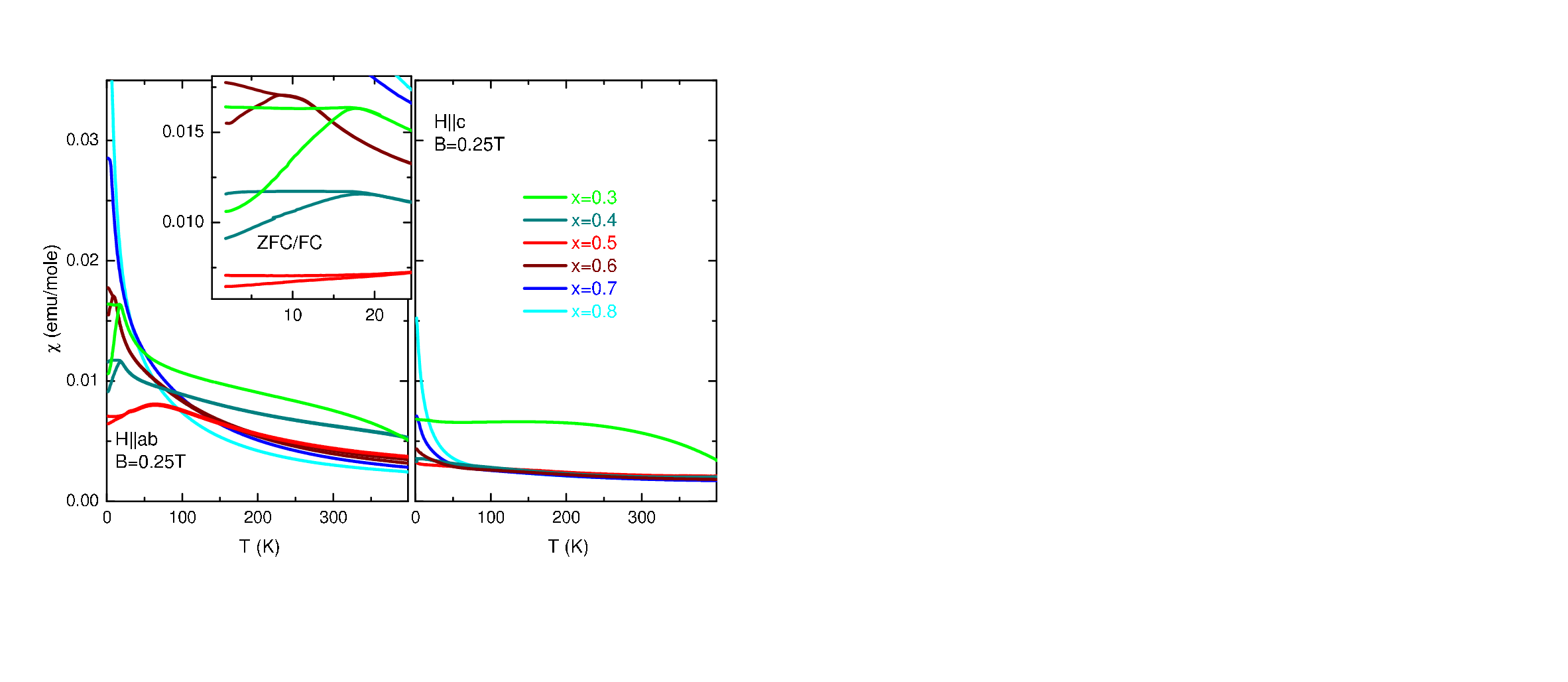}
 \caption[]{Magnetic \ch\ of \lsco\ for two different directions of the magnetic
 field. The insert is an expanded view of the low-temperature region to show the difference
between FC and ZFC measurements for $0.3\leq x \leq 0.6$.
 The curves with the lower \ch\ refer to the ZFC measurements.} \label{fig:chi}
\end{figure}

The magnetization was measured with a Quantum Design vibrating
sample magnetometer (VSM). The field was aligned parallel to the CoO$_2$ planes
as well as perpendicular to these
planes (the crystallographic $c$ direction). The corresponding components \ca\ and \cc\ are plotted in
Fig.~\ref{fig:chi}. A short-range antiferromagnetic order has been
found in the compounds $0.3\leq x\leq 0.6$ by neutron measurements\cite{cwikunp}. This
frustrated short-range order is also reflected by the difference
between field cooled (FC) and zero-field cooled (ZFC) measurements at low temperatures. The \ch\ is
smaller for ZFC than FC below a certain freezing temperature.

\begin{figure}[t]
\includegraphics[angle=0,width=\textwidth]{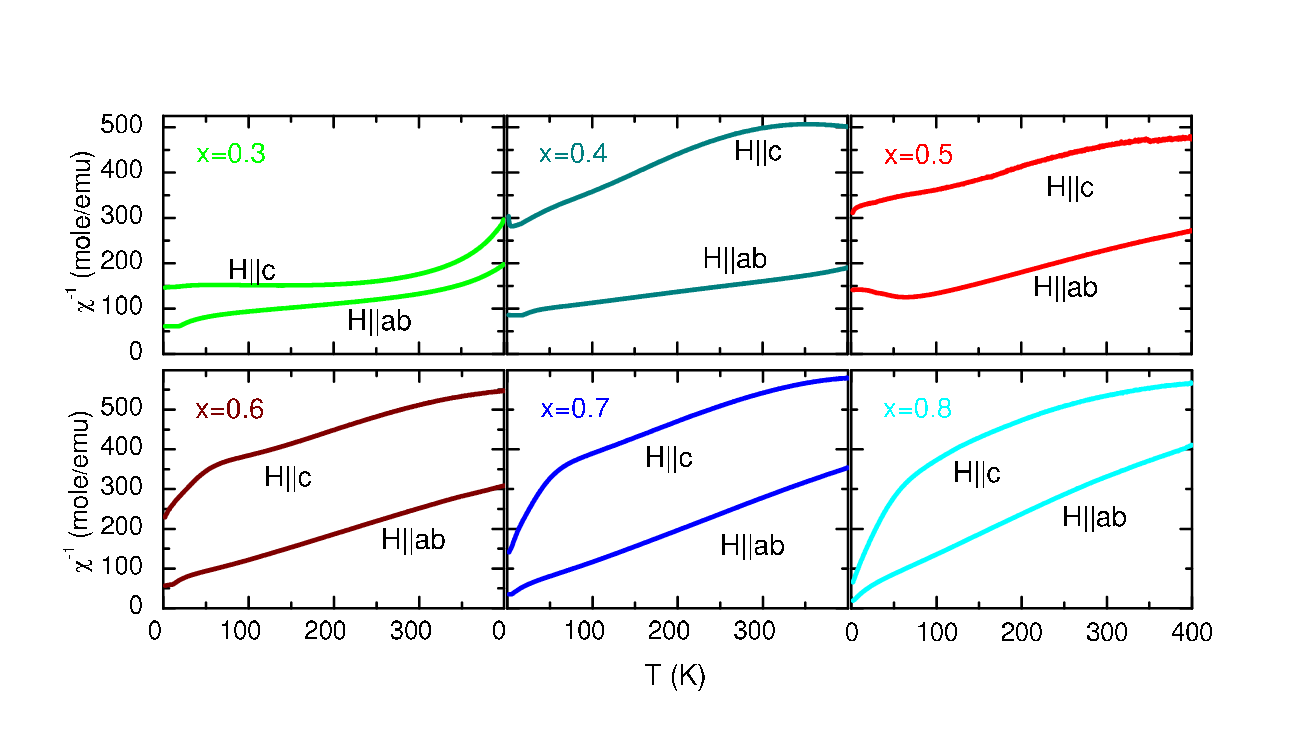}
 \caption[]{Inverse \ch\ of \lsco\ for two different directions of the magnetic
 field. The form of the curves and the magnetic anisotropy strongly deviates from Curie-Weiss behavior.} \label{fig:sus}
\end{figure}

Figure \ref{fig:sus} shows the inverse magnetic \ch\ for both
orientations of field. The main feature of the \ch\ in the paramagnetic regime is the
pronounced magnetic anisotropy, both in magnitude
and form of the curves. The direction of the anisotropy is the
same for all crystals, finding \ca\ to be bigger than \cc.

Here, the magnetic anisotropy arises from band structure
and spin-orbit coupling. For Mott and charge-transfer insulators
with well-localized moments, band-structure and
covalency effects can be approximated by an effective crystal field. The crystal field
reflects the anisotropy of the lattice and lifts the degeneracy of
the $3d$ states, resulting in a new set of linear combinations of
the unperturbed wave functions as a basis. The expectation values
of the components of the orbital moment result from these
new linear combinations. Thus, the crystal's anisotropy may result
in an anisotropy of the orbital moment. Spin-orbit coupling ties the
spin moment to this anisotropy. The spin-orbit coupling Hamiltonian will be written
as $\zeta\sum_i l_i\cdot s_i$ where the sum over $i$ runs over all electrons. The dot
product between spin and orbital momentum tends to align these moments antiparallel for each electron. For
an anisotropic orbital momentum, the spin is aligned in the direction of maximum
orbital momentum\cite{bruno89}.

The full many-body ground-state for a $d^6$ or $d^7$ configuration in a crystal-field calculation including spin-orbit
coupling and a tetragonal distortions is not simple \cite{Goodenough68} but well known. 
In order to get an intuitive picture one would like to fall back to a single electron description. In the limit of full 
spin polarization this can be done and gives important results.
In the following we will first discuss the magnetic anisotropy of La$_{2-x}$Sr$_{x}$CoO$_{4}$ in terms of a one-electron picture. 
We will show that each spin state has a different magnetic anisotropy from which, by comparison to the experiment,
the spin states of the Co ion can be concluded. In order to obtain also a quantitative description and to verify
our simple argumentation we will present a full many-body crystal-field calculation afterwards.

Within a cubic crystal structure, the $3d$ states split into
\eg\ orbitals and \tg\ orbitals. The basis can be chosen as $\{
3z^2-r^2,x^2-y^2\}$ and $\{ xy,xz,yz\}$, respectively. A partially
filled \tg\ shell can produce a pseudo orbital moment of
$\tilde{L}=1$. Though the individual real wave functions $d_{xy}$, $d_{yz}$
and $d_{xz}$ of the basis themselves have a completely
quenched orbital moment, their linear combinations are in
general complex. Writing $d^x_{m_l}$, $d^y_{m_l}$ and $d^z_{m_l}$
as the orbital wave function with the orbital moment quantized
along the axes $x$, $y$ and $z$, respectively, one finds

\begin{eqnarray}
d^x_{\pm 1}&=&\frac{1}{\sqrt{2}}(\pm d_{xy}+id_{xz})\label{eq:co21}\\
d^y_{\pm 1}&=&\frac{1}{\sqrt{2}}(\pm d_{yz}+id_{xy})\label{eq:co22}\\
 d^z_{\pm 1}&=&\frac{1}{\sqrt{2}}(\pm d_{xz}+id_{yz}).\label{eq:co3}
\end{eqnarray}

In the limit of very large crystal field splittings
$10Dq$ and with a partially filled \tg\ shell, the moment is
isotropic, despite the large orbital moment.
In fact, the \tg\ electrons are sometimes compared to $p$
electrons\cite{kamimura56a}.

\begin{figure}[t]
\begin{center}
\includegraphics[angle=270,width=10cm]{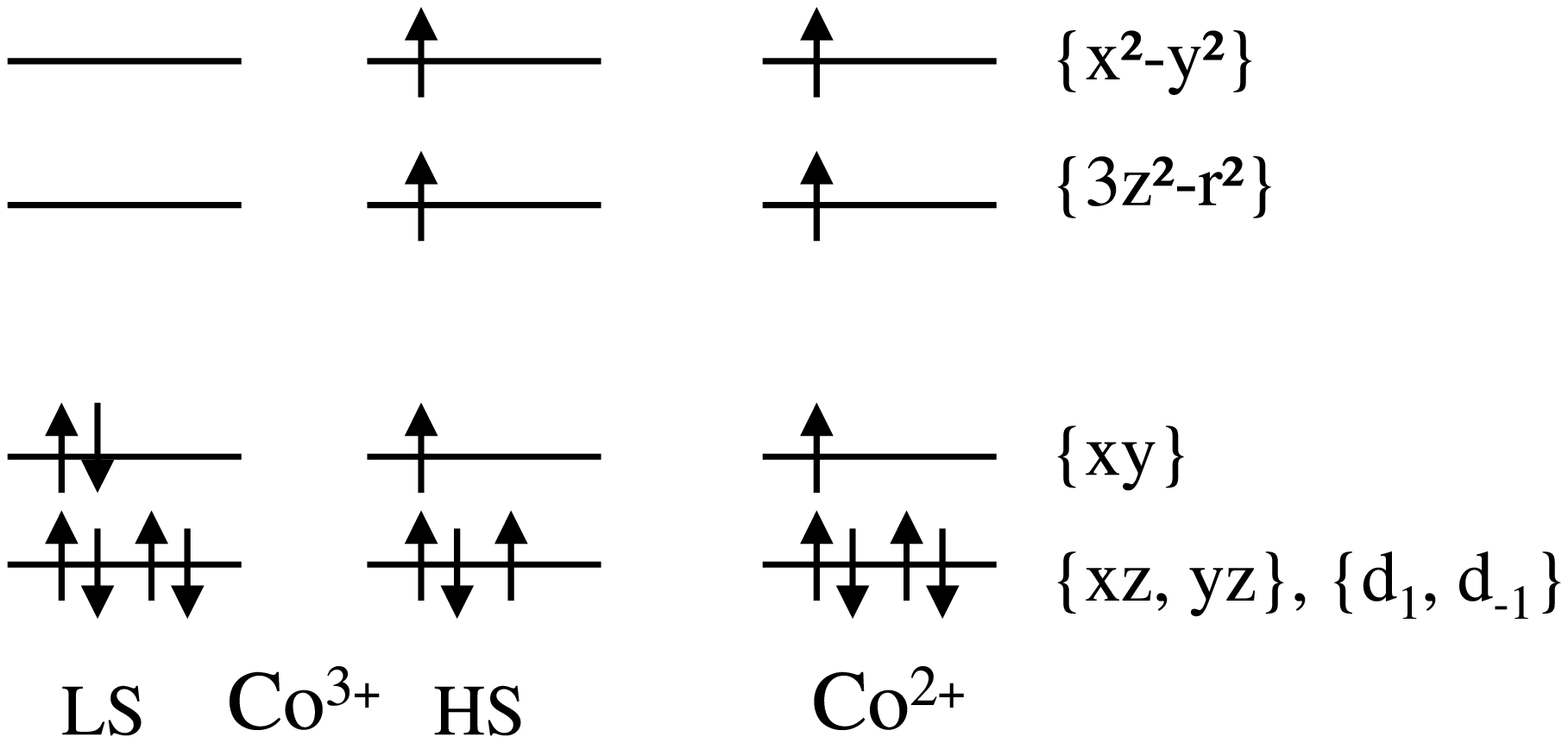}
\end{center}
 \caption[]{Splitting of the $3d$ levels in a tetragonal crystal field arising from an elongated oxygen octahedron. The two sketches on the left show the
 occupation for the LS and HS state of Co$^{3+}$ in a one-electron picture. The right
 sketch refers to HS Co$^{2+}$. The real orbitals on the right can be used as a basis.} \label{fig:occ}
\end{figure}

Introducing a tetragonal distortion, the
orbital moment becomes anisotropic. The tetragonal crystal field splits
the cubic \eg\ states into non-degenerate $a_{1g}$ and $b_{1g}$
levels, while the \tg\ states split into a non-degenerate $b_{2g}$
state and a two-fold degenerate $e_g$ level. As a basis for these
states, the real orbital functions can also be used. The $z$ axis
of the system is taken to be identical with the $c$ axis of the crystal.
In the case of \lsco, the oxygen octahedron is elongated in the
\cd. The order of levels and occupations for the
ground states of the two cobalt ions is illustrated in
Fig.~\ref{fig:occ}.

The Co$^{3+}$ low-spin state is, except for a small Van Vleck \ch,
nonmagnetic since it does not carry any moment. In the high-spin
state, Co$^{3+}$ has spin and orbital moment. The orbital degeneracy is not
completely quenched: the degenerate $xz$ and $yz$ are occupied by three electrons.
These real orbitals can thus be recombined to form complex orbitals carrying orbital moment in
the $z$ direction as described in Eq.~(\ref{eq:co3}). Note that the linear combinations
in equations (\ref{eq:co21}) and (\ref{eq:co22}) cannot be formed because the $xy$ orbital is not degenerate with the $xz$ and $yz$ orbitals. Therefore the orbital
moment is larger in the $z$ direction making this the easy axis.
This anisotropy should be reflected in the \ch\ with \cc$>$\ca, which obviously contradicts the
anisotropy found in the measurements. A Co$^{3+}$ HS system shows
the wrong magnetic anisotropy.

\begin{figure}[t]
\begin{center}
\includegraphics[angle=270,width=10.7cm]{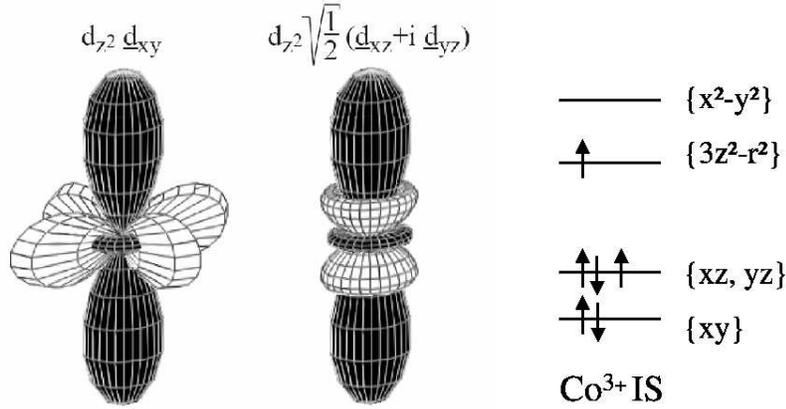}
\end{center}
 \caption[]{The picture on the left shows the charge distribution for an electron in the $3z^2-r^2$ orbital (black) and a hole in the $xy$ orbital (white). The charge distribution in the center refers to an electron in the $3z^2+r^2$ orbital and a hole in a linear combination of the $xz$ and $yz$ orbitals. The splitting and the occupation of the $3d$ levels for Co$^{3+}$ in the IS state is shown on the right.} \label{fig:is}
\end{figure}

Next, we discuss the IS state of Co$^{3+}$. Due to the elongation of the oxygen octahedra, the \eg\ electron occupies the $3z^2-r^2$ orbital. This also effects the splitting of the \tg\ states.
The $xz$ and $yz$ orbitals remain degenerate but are not degenerate with the $xy$ orbital. This arises from the different charge distributions in relation to the $3z^2-r^2$ orbital.
We have five electrons to fill in the \tg\ states, which can also be treated as one hole in the
\tg\ states. This hole is attracted to the \eg\ electron in the $3z^2-r^2$ orbital. Fig.~\ref{fig:is} shows the charge distribution of this hole and the $3z^2-r^2$ orbital. In the left sketch, both the electron in the $3z^2-r^2$ orbital and a hole in the $xy$ orbital is drawn. The sketch in the center shows the $3z^2-r^2$ orbital and the hole in a linear combination of the degenerate $xz$ and $yz$ orbitals. Regarding the distances between the electron and the hole, on can conclude that the state with the hole in the $xz$ and $yz$ orbitals is lower in energy. As the order of levels is reversed when we are speaking about holes instead of electrons, this means that the $xy$ orbital is lowered in energy.
Thus, the order and occupation of levels is the one shown in Fig.~\ref{fig:is}. The magnetic anisotropy is the same as in the case of Co$^{3+}$ HS which we treated in the last paragraph. With three electrons in the degenerate level of the $xz$ and $yz$ orbitals, we can also make use of Eq.~(\ref{eq:co3}) to show that the direction of anisotropy does not fit the measurements for \lsco.

Neither the HS nor the IS state of Co$^{3+}$ show the correct magnetic anisotropy. But before we draw further conclusions,
Co$^{2+}$ should be discussed. Although Co$^{2+}$ has also been found
in the LS state in some intramolecular compounds\cite{brooker02}, for bulk crystals
it can be safely assumed that Co$^{2+}$ is in the HS state\cite{goodenough63}.

Regarding the ground state produced by the crystal field in Fig.~\ref{fig:occ}, where the lowest
level is filled, no linear combinations like in Eq.~(\ref{eq:co3})
can be formed. The orbital moment would be completely quenched
and the magnetic moment would be determined by an isotropic spin.
If, however, spin-orbit coupling is of the same magnitude as the
splitting of the \tg\ orbitals, it will mix these states.
Mixing in a state with orbital moment in the $z$ direction would require placing a hole in
the $d_{\pm 1}^z$ orbital (see Eq.~\ref{eq:co3}). This would cost an energy equal to the splitting within the
\tg\ orbitals. It is more favorable to mix in a state with orbital moment
in the $x$ (or $y$) direction by placing a hole in the $d_{\pm 1}^x$ (or the $d_{\pm 1}^y$)
orbital (see Eq.~\ref{eq:co21}).
This costs only half of the crystal-field splitting within the \tg\ orbitals.
An orbital moment in the $x$ (or $y$) direction costs less energy than in the
$z$ direction. So, the $xy$ plane will be the easy plane for Co$^{2+}$ in this elongated
tetragonal structure\cite{csizar05a}. The spin
moment will also be found in the $xy$ plane. Comparing these findings to the
measurement, we see that Co$^{2+}$ shows the correct direction of
magnetic anisotropy. Now a conclusion about the spin state of Co$^{3+}$ can
be drawn. For $x\ge 0.5$, at least half of the cobalt ions in the crystal
are Co$^{3+}$. Taking into account that Co$^{2+}$ has a smaller
spin moment than HS Co$^{3+}$, the latter would dominate the
anisotropy. The opposite is, however, the case in the data. Thus, for this doping range, the
spin state of Co$^{3+}$ is identified with the LS state and the
magnetization must be assigned to Co$^{2+}$, which shows the
correct anisotropy. For $x=0.4$, we still have 40\% Co$^{3+}$ and,
because of the pronounced anisotropy, we can follow the same argumentation
here. Regarding $x=0.3$ in Fig.~\ref{fig:sus}, the anisotropy is
found to be rather small compared to the other compounds.

It should be stressed that the direction of anisotropy in layered perovskites depends on the direction in
which the oxygen octahedron is distorted. In some systems like K$_2$CoF$_4$ the octahedra are compressed in the
\cd , resulting in an easy-axis anisotropy\cite{folen68}; Co$^{2+}$ in almost cubic symmetry also tends to have
an easy-axis anisotropy like in CoO. However, in our case of La$_{2-x}$Sr$_x$CoO$_4$, the oxygen octahedron 
surrounding the cobalt ion is elongated by the inherent tetragonal structure. This situation is comparable to
the change of anisotropy in CoO films on different substrates\cite{csizar05a}. 

\begin{figure}[t]
\includegraphics[angle=0,width=\textwidth]{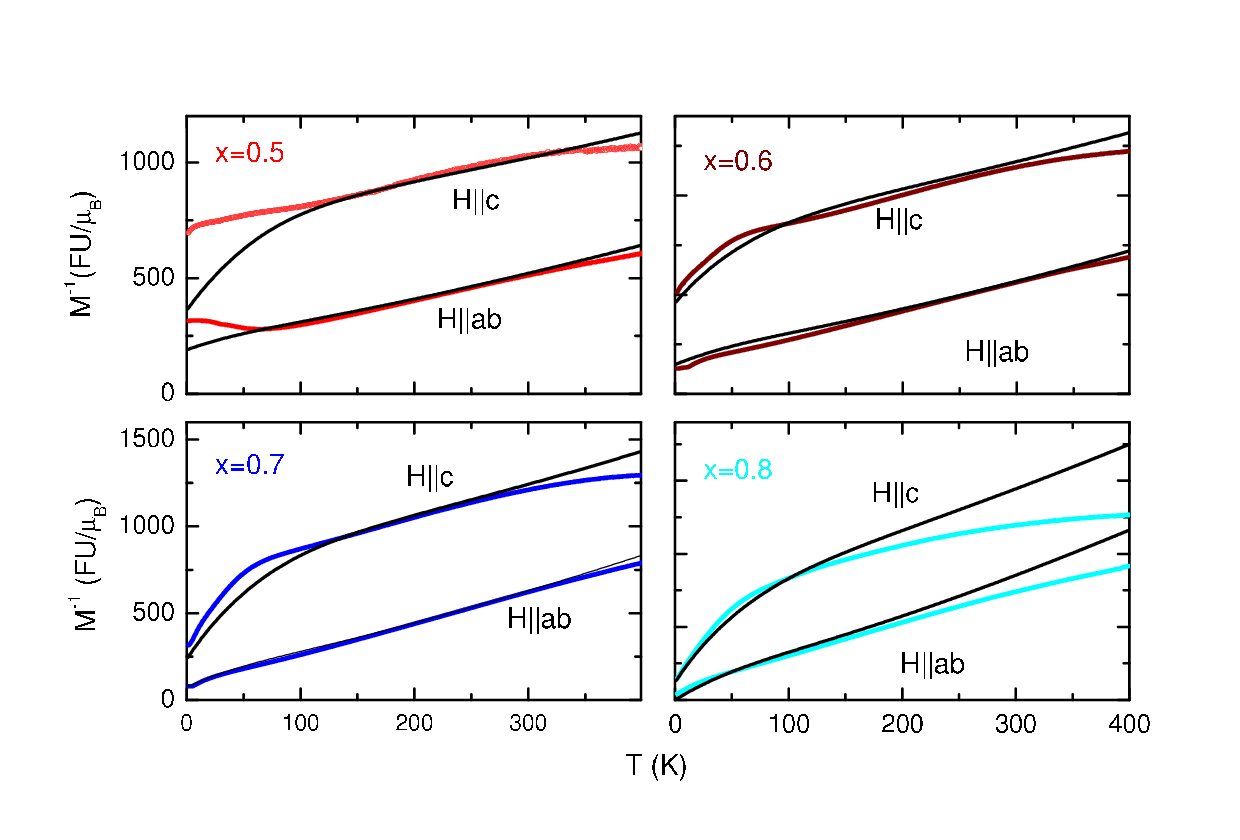}
 \caption[]{Measured magnetization (colored lines) compared
 with the calculation for Co$^{2+}$ (black lines), plotted inverse. A mean-field magnetic exchange
 was added to the calculation. Note that the magnetization of the ion was weighted
 with the content of Co$^{2+}$ ($1-x$).} \label{fig:fit}
\end{figure}

For a quantitative analysis of the data and a confirmation of the results
found in the qualitative discussion of the anisotropy, a full-multiplet
calculation was carried out within the crystal field
approximation\cite{hamil}. A Hamiltonian for the $3d$ shell of the Co$^{2+}$
ion was set up, including crystal field, spin-orbit coupling and
magnetic field. The Slater integrals $F_0$, $F_2$ and $F_4$
characterizing the Coulomb interaction and the spin-orbit coupling
constant were taken from a Hartree-Fock approximation
\cite{haverkortDiss}. Any excitations from the oxygen $2p$ levels
and mixing with higher states were neglected, as these effects are
not significant when considering thermal energies. The parameters for the
crystal field were treated semi-empirically. Regarding the
tetragonal distortion as a perturbation of the cubic symmetry, the effect of the
hybridization of the Co $3d$ shell with the surrounding oxygen
$2p$ shells only increases the splitting of the cubic \tg\ and
\eg\ states. Hybridization can thus be included into the crystal
field parameters and need not be treated separately. The
Hamiltonian was diagonalized and Eigenstates and Eigenvalues were
used to obtain the temperature dependence of the magnetization.
Magnetic exchange was included in the calculation on a mean-field
level. The results can be seen in Fig.~\ref{fig:fit}, where the inverse magnetization is plotted.

The crystal-field parameters were chosen as follows: the
$10Dq$ splitting and the splitting within the \eg\ levels
were fixed at 1.5eV and 0.2eV, respectively. These two values are
of minor importance for the calculation, because the anisotropy of
the magnetization arises from the \tg\ electrons, as
described in the discussion above. The
splitting of the \tg\ states showed up to be crucial for the form
and the anisotropy of the calculated curves.
The values of the splitting within the \tg\ states were
chosen as 80meV, 60meV, 68meV, and 50meV for the compounds $x=$0.5, 0.6, 0.7,
and 0.8, respectively.
These $t_{2g}$ crystal field splittings for Co are in good 
agreement with previous measurements on strained CoO thin films \cite{csizar05a}. 
They might appear to be
surprisingly small compared to the values found for the Titanates \cite{haverkort05Ti,rueckamp05a}
in the order of 200 meV. 
The difference between Co and Ti can be understood by considering the 
radial extent of the $d$ wave function and the transition metal - oxygen (TM-O) distance. The Co $d$ wave function is much 
smaller than the Ti $d$ wave function, reducing the covalency and thereby 
the size of the crystal-field splitting. At the same time, the average 
TM-O distances of La$_{2-x}$Sr$_{x}$CoO$_{4}$ and 
LaTiO$_{3}$ are almost equal ($\approx$2.02\AA\ for Co-O \cite{cwikunp} and $\approx$2.04\AA\ for Ti-O \cite{cwik05a}). This is related to the 
occupation of the anti-bonding $e_{g}$ orbitals in a HS Co$^{2+}$ compound 
which pushes the O atoms further away.

It can be seen in Fig.~\ref{fig:fit} that the calculation
gives a good description for $x=0.6$ and $x=0.7$. The magnetization of the half-doped
sample $x=0.5$ is reproduced well for higher temperatures. At
temperatures below the freezing temperature, the calculation cannot
describe the system because magnetic order is not included in
the model. The $x=0.8$ sample is fitted well for lower
temperatures. Above $T\approx 350$K, the magnetization in the
measurement is enhanced by another effect and results in a
deviation from the calculation. This effect can be seen in all
curves in Fig.~\ref{fig:sus}, but at significantly higher
temperatures. This increase of magnetization could arise from the thermal population of the Co$^{3+}$ HS or IS state. But
surely further measurements at higher temperatures are needed to
confirm this assumption. Nevertheless, the full-multiplet
calculation of a pure Co$^{2+}$ system already succeeds in describing the
main features of the magnetization for $x\geq 0.5$. This confirms
that Co$^{3+}$ is a LS system in this doping range.

In summary, the magnetic \ch\ of \lsco\ has been analyzed for two
different directions of the external magnetic field. A definite anisotropy of
the magnetic moment is found experimentally with \ca$>$\cc. This direction of
anisotropy does not match with the behavior of HS or IS Co$^{3+}$, whereas
Co$^{2+}$ in the HS state shows the correct single-ion anisotropy. Thus, the spin state of Co$^{3+}$ for $x\ge 0.4$
must be the LS state. This conclusion is also confirmed by a full-multiplet
calculation.

We acknowledge financial support by the Deutsche
Forschungsgemeinschaft through SFB\,608.

\section*{References}
\bibliographystyle{unsrt}

\end{document}